\documentclass[twocolumn,groupedaddress,assymb,amsmath]{revtex4}
\usepackage{graphicx}
\usepackage{amssymb}
\usepackage{amsmath}

\usepackage{extarrows}
\usepackage{color}
\usepackage[colorlinks]{hyperref}

\begin{document}


\title{Bloch band structures and linear response theory of  nonlinear systems}

\author{Fude Li$^{1}$, Junjie Wang$^{1}$, Dianzhen Cui$^{1}$, K. Xue$^{1}$, and X. X. Yi$^{1}$}
\email{yixx@nenu.edu.cn}

\affiliation{$^1$Center for Quantum Sciences and School of Physics, Northeast Normal University, Changchun 130024, China}

\date{\today}

\begin{abstract}
We investigate the Bloch bands and develop a linear response theory for  nonlinear systems, where the interplay between topological parameters and nonlinearity leads to new band structures. The nonlinear system under consideration is described by the Qi-Wu-Zhang model with Kerr-type nonlinearity, which can be treated as a nonlinear version of Chern insulator. We explore the eigenenergies of the  Hamiltonian and discuss its Bloch band structures as well as the condition of gap closing. A cone structure in the ground Bloch band and tubed structure in the excited Bloch band is found.  We also numerically calculate the linear response of the nonlinear Chern insulator to external fields, finding that these new band structures   break the condition of adiabatic evolution and make the linear response not quantized. This feature of  response  can be understood  by examining  the dynamics of the nonlinear system.
\end{abstract}

\keywords{Linear response in nonlinear system; Chern model}
\maketitle


\section{introduction}\label{sec1}
In the last decades, the rapid development of topological band theory in condensed matter physics\cite{Thouless1982,klitzing1986,Kane2010,qi2011,Ando2013} has found applications in many   materials, such as topological insulators\cite{Zhang2006,Zhang2007,Hsieh2008,
Hsieh2009,moore2011,Kane2007,Mele2007},   topological semimetals\cite{Sergey2011,Ding2015,Xu2015,Chen2014}, and topological crystalline insulators\cite{Fu2011,Jia2014,Dziawa2012,Hsieh2012,Tanaka2012}. Topology of energy bands structures of these materials offer us a new method to classify  different quantum phases with topological invariant\cite{Thouless1982,Berry1984,Kohmoto1985,Niu2010,Zak1989} and lead to many fascinating phenomena including topological pumps and quantized conductance.   The linear response of topological materials to external fields can be described  by topological invariants. For example,  quantized Hall conductance is proportional to Chern numbers, which can be verified  by the linear response of these electronic systems to  external electronic fields.

Recently, topological concepts developed for electric systems have been  extended to study a variety of photonic fields\cite{Lu2014,Ozawa2019,Khanikaev2013,Khanikaev2017,Chen2017,Carusotto2016} such as photonic crystals\cite{Hu2015,Haldane2008,tikhodeev2012}, waveguide arrays\cite{Raghu2008,shvets2015,Rechtsman2013,silberberg2013,Zhang2021} and ring resonators\cite{yuan2018,Hafezi2011,Hafezi2013}. These systems possess topological band structure similar to the electric systems. In this sense, topological band structures can be  realized (or simulated) and observed in photonic systems\cite{polman2019,Angelakis2021,Rechtsman2018}. Due to the manufacturing flexibility, optical systems may offer more modulating parameters than their electric counterparts including non-linear couplings.

Combining these topological photonic structures with nonlinear effects, we will make a step forward in  the emerging field of nonlinear topological photonics, which would fill the gap between the physics of topological phases and  nonlinear optics\cite{kivshar2020,kinsey2022,
mluckov2021,Ye2019,Liew2015,hadad2016,christodoulides1988,
qi2012,mork2015,fleury2019,thecharis2021,raditya2021,leykam2021,Tang2021,Yuri2019}.
In fact, there are many publications in this field for two- or many-level systems. For example, looped structures are produced  in one-dimensional systems\cite{niu2000,niu2002,wu2002,liu2000,smith2003,niu2003,liu2007,wang2008,Chen2019,tuloup2020,konotop2021}, which have recently been realized experimentally\cite{porto2016,blume2020,alexander2021,Hadad2018}. For Bloch band in nonlinear systems, interplay between nonlinearity and  band topology  can lead to many interesting structures.
Nonlinearities arising from photonic structures have an effect on topological phases\cite{yao2022,georgios2021,Delplace2022,charalampos2022} and trigger the appearance of new band structures\cite{wu2011,gong2017,smirnova2021,Yi2021,He2020,zhu2022}, such as nonlinear Dirac cones and nonlinear Dirac points  in two-dimensional systems.
Generally, linear topological bands can be combined with mean-field interactions, which reduces any many-body problem into an one-body nonlinear problem. In the mean-field approximation, the discussions on the nonlinear systems are based on nonlinear Schr\"odinger equation with state-dependent Hamiltonian(nonlinear Hamiltonian), which is widely used to study emerging nonlinear features.

Except the study of the effect of nonlinearity on Bloch bands, here we also focus on the linear response of nonlinear systems to external field which plays a prominent role in many fields of physics. Intersection between linear topological bands and nonlinearity can give rise to the radical change of linear response of  mean-field nonlinear systems.
In contrast to the linear response in linear systems,   the linear response of nonlinear systems to external field is not quantized and can  not be  characterized  by topological invariant.

The reminder of this manuscript is organized as follows.
In Sec.~\ref{sec2}, we develop a method to explore the Bloch band structures in a general nonlinear two-band systems.
In Sec.~\ref{sec3}, we illustrate the method developed in the last section with  a nonlinear Chern insulator. Band gap closing of the system  is also discussed.
In Sec.~\ref{sec4}, we numerically calculate  the linear response of nonlinear Chern insulator to external field and discuss the  differences of linear response in linear and nonlinear systems.
We consider adiabatic  evolution of the system governed by the nonlinear Bloch Hamiltonian and study effects brought by the dynamics beyond the adiabatic evolution.
In Sec.~\ref{sec5}, we examine  the dynamics of the  system to study whether the adiabatic evolution has an effect on the linear response of the system.
In Sec.~\ref{sec6}, we summarize our results.

\section{Bloch band structure for nonlinear systems}\label{sec2}
We begin by presenting  a method to find the eigenvalues of general nonlinear systems. To be specific, we consider a two-band system whose dynamics being  governed by the following nonlinear Schr\"odinger equation
\begin{eqnarray}\label{}
i\partial_t\Psi(t)=\hat{H}(t)\Psi(t)
\end{eqnarray}
with
\begin{eqnarray}\label{}
\hat{H}(t)&=&\mathbf{d}\cdot\hat{\sigma}
+ U[|\psi_1(t)|^2,0;0,|\psi_2(t)|^2],
\end{eqnarray}
where $\hat{\mathbf{\sigma}}=(\sigma_x,\sigma_y,\sigma_z)$ are Pauli matrices acting on the pseudospin space, and $\mathbf{d}=(d_x,d_y,d_z)$. This Hamiltonian can be obtained by considering the Bloch form of a two-dimensional Bose-Hubbard  Hamiltonian by the mean-field approximation in the limit of the large number of particles. And $\Psi=[\psi_1(t),\psi_2(t)]^T$ denotes the time-dependent mean-field wave function. We consider the  time-independent form $\psi_j=\phi_j e^{-i\epsilon t/\hbar},j=1,2$,  and define the nonlinear Hamiltonian  as
\begin{eqnarray}
\hat{H}=\left(
\begin{array}{cc}
d_z+ U|\phi_1|^2& d_x-id_y \\
d_x+id_y & -d_z+U|\phi_2|^2
\end{array}\right).
\end{eqnarray}
The time-independent nonlinear eigenvalues  and corresponding eigenstates can be found by solving
\begin{eqnarray}\label{}
\hat{H}
\left(
\begin{array}{c}
   \phi_1 \\
   \phi_2
\end{array}\right)
=\epsilon
\left(
\begin{array}{c}
   \phi_1 \\
   \phi_2
\end{array}\right).
\end{eqnarray}
Here these  eigenstates correspond to stationary solutions of time-dependent nonlinear Schr\"odinger equation. Due to nonlinearity, more eigenenergies might appear in the system. This is one of the key differences between linear and nonlinear systems.   To solve this eigenvalue equation, we set a new variable $\kappa$, $\kappa = |\phi_1|^2-|\phi_2|^2$, and use the normalization condition, $|\phi_1|^2+|\phi_2|^2=1$.
After tedious but straightforward  calculations, the relation between $\kappa$ and $\epsilon$ is given by
\begin{eqnarray}
(\epsilon-U)\kappa=d_z,
\end{eqnarray}
and the eigenstates, $|\phi_1|^2$ and $|\phi_2|^2$, are given by
\begin{eqnarray}\label{}
 |\phi_1|^2&=&\frac{1}{2}+\frac{d_z}{2(\epsilon-U)},\nonumber \\
 |\phi_2|^2&=&\frac{1}{2}-\frac{d_z}{2(\epsilon-U)},
\end{eqnarray}
where we replace $\kappa$ with $\epsilon$,  and  get an $\epsilon$-dependent nonlinear Hamiltonian useful for our calculations. In linear regime, the eigenvalues can be obtained from $|H-\epsilon|=0.$ In nonlinear regime, this condition is  the same as in the linear regime. Setting the determinant be zero,
\begin{eqnarray}
\left|
\begin{array}{cc}
 d_z+\frac{U}{2}+\frac{U}{2}\kappa-\epsilon & d_x-id_y  \\
 d_x+id_y & -d_z+\frac{U}{2}-\frac{U}{2}\kappa-\epsilon
\end{array}
\right|=0,
\end{eqnarray}
we obtain
\begin{eqnarray}\label{energy}
f(\epsilon) = \epsilon^4&-&3 U \epsilon^3+(\frac{13}{4}U^2-d_z^2-d_x^2-d_y^2)\epsilon^2\nonumber\\
&+&(U d_z^2+2 U (d_x^2+d_y^2) -\frac{3}{2} U^3 ) \epsilon\nonumber\\
&+&\frac{U^4}{4}-\frac{U^2d_z^2}{4}-U^2(d_x^2+d_y^2)=0,
\end{eqnarray}
Solving this equation, we can find three  different types of degenerate eigenvalues called degenerate points. The I-type degenerate point is $\epsilon=U/2$ when $d_x^2+d_y^2=0$, the II-type degenerate point is $\epsilon=U$ when $d_z=0$, and  the III-type degenerate point can be obtained by setting $d_z=\pm\frac{1}{2}\{U^{\frac{2}{3}}-[4(d_x^2+d_y^2)]^{\frac{1}{3}}\}^{\frac{3}{2}}$.  $f(\epsilon)=0$ leads to,
\begin{eqnarray}
\{2(\epsilon-\frac{U}{2})&-&[4U(d_x^2+d_y^2)]^{\frac{1}{3}}\}^2\{4(\epsilon-\frac{U}{2})^2\nonumber\\
&+&[-4U+4[4U(d_x^2+d_y^2)]^{\frac{1}{3}}](\epsilon-\frac{U}{2})\nonumber\\
&-&U[4U(d_x^2+d_y^2)]^{\frac{1}{3}}\}=0,
\end{eqnarray}
where  the III-type degenerate point is at
\begin{eqnarray}
\epsilon =
\frac{U}{2} + \frac{1}{2} [4U(d_x^2+d_y^2)]^{\frac{1}{3}}.
\end{eqnarray}
In the following, we would examine these  eigenvalues and show that bifurcation appears around these degenerate points. This means that  additional eigenvalues in nonlinear Hamiltonian would appear. This is why nonlinear systems possess  new Bloch band structures that are completely different from their linear counterparts.

Before proceeding to analyze the three types of degenerate eigenvalues,
we approximate nonlinear Bloch Hamiltonian around the degenerate points by using Taylor expansions, which leads to a correction of  $\epsilon^{(1)}$ to the eigenvalue $\epsilon^{(0)}$   up to the first order in the deviation from the degenerate points.  The first-order correction  $\epsilon^{(1)}$ is given by
\begin{eqnarray}
a(\epsilon^{(1)})^2+b \epsilon^{(1)}+c=0,
\end{eqnarray}
where
\begin{eqnarray}
a&=&\frac{1}{2}(\frac{\partial^2 f(\epsilon) }{\partial \epsilon^2})^{(0)},\nonumber\\
b&=&(\frac{\partial f(\epsilon)}{\partial \epsilon})^{(0)}+(\frac{\partial^2 f(\epsilon) }{\partial \epsilon\partial d_\mu})^{(0)} d_\mu^{(1)},\nonumber\\
c&=&(\frac{\partial f(\epsilon)}{\partial d_\mu})^{(0)} d_\mu^{(1)}
+\frac{1}{2}(\frac{\partial^2 f(\epsilon) }{\partial d_\nu\partial d_\mu})^{(0)}d_\nu^{(1)}d_\mu^{(1)}.
\end{eqnarray}
Here $()^{(0),(1)}$ denotes the zero-order or first-order values, which is parameter dependent. Generally, this parameter is wavevector $\mathbf{k}
$ in the momentum space, thus the zero-order or first-order values at $\mathbf{k}=\mathbf{k}_0$ can be given by $d_\mu^{(0)}=d_\mu(\mathbf{k})|_{\mathbf{k}=\mathbf{k}_0}, d_\mu^{(1)}=(\mathbf{k}-\mathbf{k}_0)\cdot\partial_{\mathbf{k}}d_\mu(\mathbf{k})|_{\mathbf{k}=\mathbf{k}_0}, \mu=x,y,z$ and $f(\epsilon)$ is given in  Eq.~(\ref{energy}).
The values of $b^2-4ac$  can be regarded as a critical condition for the appearance of bifurcation.

Consider the I-type degenerate point $\epsilon^{(0)}=U/2$ and $d_x^{(0)}=0,d_y^{(0)}=0$,  $\epsilon^{(1)}$ is  given by
\begin{eqnarray}
\epsilon^{(1)}=\pm\frac{U\sqrt{(d_x^{(1)})^2+(d_y^{(1)})^2}}{\sqrt{U^2-4(d_z^{(0)})^2}}.
\end{eqnarray}
From the last equation, it is easy to find that  if $U_c\neq 2|d_z^{(0)}|$, $\epsilon^{(1)}$ would get two values, then  $U_c=2|d_z^{(0)}|$ can be treated as the critical condition for  bifurcation.

Around the II-type degenerate point $\epsilon^{(0)}=U$ with $d_z^{(0)}=0$, $\epsilon^{(1)}$ is given by
\begin{eqnarray}
\epsilon^{(1)}=\pm\frac{U}{2\sqrt{2}\sqrt{U^2-4((d_x^{(0)})^2+(d_y^{(0)})^2)}}.
\end{eqnarray}
Similarly,  $U_c=2\sqrt{(d_x^{(0)})^2+(d_y^{(0)})^2}$ is the  critical condition for  bifurcation. Critically, the III-type degenerate point is connected to  the closing of gap between different Bloch bands, we will perform  detailed discussion in the following sections.


\section{Bloch band structures in nonlinear Chern insulator}\label{sec3}
In this section, we analyze Bloch band structures in momentum space focusing   our attention on Bloch form of nonlinear Hamiltonians. We introduce a simple nonlinear system by adding the Kerr-type nonlinearity into the Chern insulator model. The dynamics of such a system  in terms of Bloch form of wave eigenfunctions is governed by
\begin{eqnarray}\label{timenl}
i\partial_t\Psi(\mathbf{k},t)=\hat{H}(\mathbf{k},t)\Psi(\mathbf{k},t),
\end{eqnarray}
where the wave vector $\mathbf{k}=(k_x,k_y)$ is restricted to $k_{x,y}\in[0,2\pi]$ called  the restricted Brillouin zone. Consider a two-band model and write $\Psi(\mathbf{k},t)=[\psi_1(\mathbf{k},t),\psi_2(\mathbf{k},t)]^T$  and $\hat{H}(\mathbf{k},t)=\hat{H}_L(\mathbf{k})+\hat{H}_{NL}(\mathbf{k},t)$. $\hat{H}_L(\mathbf{k})$ and $\hat{H}_{NL}(\mathbf{k})$ are linear and nonlinear parts of the Bloch Hamiltonian
\begin{eqnarray}\label{nlchern}
\hat{H}_L(\mathbf{k})&=&d_x(\mathbf{k})\sigma_x+d_y(\mathbf{k})\sigma_y+d_z(\mathbf{k})\sigma_z,\nonumber\\
\hat{H}_{NL}(\mathbf{k})&=&U[|\psi_1(\mathbf{k},t)|^2,0;0,|\psi_2(\mathbf{k},t)|^2],
\end{eqnarray}
with
\begin{eqnarray}
\mathbf{d}(\mathbf{k})&=&(d_x(\mathbf{k}),d_y(\mathbf{k}),d_z(\mathbf{k}))\nonumber\\
&=&(J\sin{k_x},J\sin{k_y},u+J\cos{k_x}+J\cos{k_y}),\nonumber\\
\end{eqnarray}
where $u$ is a topological parameter that can change the topological phases and $U$ is nonlinear strength. Throughout this manuscript, we set $J=1$ and take $J$ as the units of energy and $1/J$ as the units of time.

In order  to obtain Bloch band structures,  we consider the stationary solution $\Psi(\mathbf{k},t)=e^{-i\epsilon t}\Phi(\mathbf{k})$ to the Schr\"odinger equation. The time-independent eigenvalue equation is given by
\begin{eqnarray}\label{eigennl}
\hat{H}(\mathbf{k})\Phi(\mathbf{k})=\epsilon\Phi(\mathbf{k}).
\end{eqnarray}

\subsection{Linear regime}
We first consider Bloch band structures in linear Bloch Hamiltonian. When $U=0$, Bloch Hamiltonian $\hat{H}(\mathbf{k})$ is $\hat{H}_L(\mathbf{k})$ and eigenvalues of $\hat{H}_L(\mathbf{k})$ are
\begin{eqnarray*}
\epsilon_{\pm} &=& \pm\sqrt{u^2+2+2u\cos{k_x}+2u\cos{k_y}+2\cos{k_x}\cos{k_y}}.
\end{eqnarray*}
There is an energy gap between these two Bloch bands. And the energy gap is closed when $u=0,\pm2$ at the special points in $\mathbf{k}$ space.  These points are  the so-called Dirac points. Since we consider linear Bloch bands, we can characterize different topology of band by Chern number defined as $C=\frac{1}{2}sgn(u+2)+\frac{1}{2}sgn(u-2)-sgn(u)$. Thus, different values of $u$ would correspond to different topology of Bloch bands\cite{Bernevig2013}.

\begin{figure}
\includegraphics[width=0.5\textwidth]{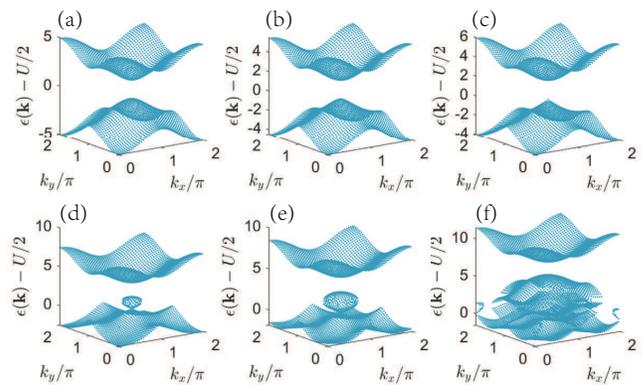}
\caption{Bloch band structures in the topological trivial regime $u=3$ with different  $U$ as in (a) $U=0$, (b) $U=1$, (c) $U=2$, (d) $U=5$, (e) $U=7$, and (f) $U=13$.}\label{energy1}
\end{figure}

\subsection{Nonlinear regime}
Now, we turn to the nonlinear Bloch Hamiltonian. Unlike linear Bloch Hamiltonian, we can not use Chern number to distinguish different topological phases due to appearance of degenerate points in Bloch band. Nevertheless, we can still continue to discuss Bloch band structures because they are different from linear one.  With Eq.~(\ref{energy}) and  some elaborate algebra, a $\mathbf{k}$-dependent algebraic equation for eigenvalues $\epsilon$ can be given
\begin{eqnarray}
\epsilon^4(\mathbf{k})&-&3 U \epsilon^3(\mathbf{k})+[\frac{13}{4}U^2-d_z^2(\mathbf{k})-d_x^2(\mathbf{k})-d_y^2(\mathbf{k})]\epsilon^2(\mathbf{k})\nonumber\\
&+&[U d_z^2(\mathbf{k})+2 U (d_x^2(\mathbf{k})+d_y^2(\mathbf{k})) -\frac{3}{2} U^3] \epsilon(\mathbf{k})\nonumber\\
&+&\frac{U^4}{4}-\frac{U^2d_z^2(\mathbf{k})}{4}-U^2(d_x^2(\mathbf{k})+d_y^2(\mathbf{k}))=0.
\end{eqnarray}
In the following, we tune nonlinear strength $U$ and topological parameter $u$ to show different band structures.

We first consider topological trivial case, i.e. $u>2$. In Fig.~\ref{energy1}, we plot several Bloch band structures with different nonlinear strength $U$ and fixed $u=3$ as well as $d_x(\mathbf{k})=d_y(\mathbf{k})=0$.   The I-type degenerate points can be easily found. Using the critical condition of the I-type degenerate point given in the last section, we obtain the critical nonlinear strength,   $U_c=2|u+2|,2|u|,2|u|,2|u-2|$. For $u=3$ in the trivial regime, critical nonlinear strength are $U_c=10,6,6,2$. When $U<U_c$, as shown in Fig.~\ref{energy1}(a)-(c),  with $U$ increasing, Bloch band structures start to change, see Fig.~\ref{energy1}(d). Cone structures can be formed, which are regarded as nonlinear Dirac cones\cite{gong2017}. Tap between different Bloch bands exists permanently with increasing $U$,  which is different  from the topological nontrivial regime, see   Fig.~\ref{energy1}(f). It is worth noting that in topological trivial regime, we cannot obtain the II-type degenerate points because the condition $d_z(\mathbf{k})=u+\cos{k_x}+\cos{k_y}=0$ cannot be satisfied.

\begin{figure}
\includegraphics[width=0.5\textwidth]{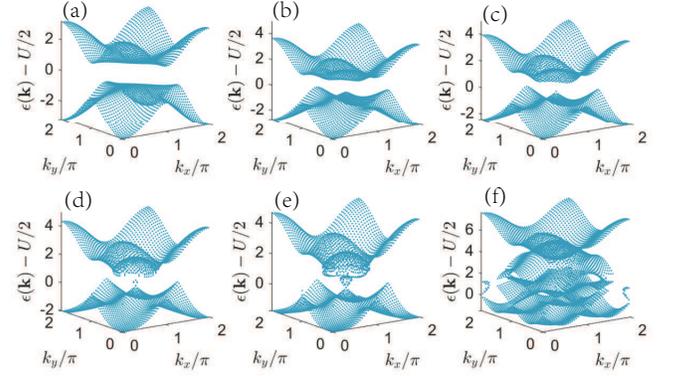}
\caption{Bloch band structures in topological nontrivial regime $u=1.2$. With increasing $U$ in (a) $U=0$, (b) $U=1$, (c) $U=1.6$, (d) $U=2.4$, (e) $U=3$, (f) $U=9$.}\label{energy2}
\end{figure}

In topological nontrivial regime, we set the  topological parameter $u=1.2$ and $U=0,1,1.6,2.4,3,9$ was chosen to discuss the problem. In this case,  $U_c=2|u+2|,2|u|,2|u|,2|u-2|=6.4,2.4,2.4,1.6$. By $u+\cos{k_{x_0}}+\cos{k_{y_0}}=0$ and $d_z(\mathbf{k})=0$, we can obtain the corresponding II-type degenerate points in the upper band. Using the critical condition of the II-type degenerate point in the last section, we can give the corresponding critical nonlinear strength $U_c=2\sqrt{(\sin{k_{x_0}})^2+(\sin{k_{y_0}})^2}$. When $U<U_c$, as shown in Fig.~\ref{energy2}(a)-(c), Bloch band structure remain unchanged. As $U$ increases, eigenvalues in the vicinity of the I-type degenerate points in the lower band become sharper, see Fig.~\ref{energy2}(c).  The other cone structures in the lower band and tubed structures in the upper band can be found in Fig.~\ref{energy2}(d)-(e). And gap between different Bloch bands disappears with increasing $U$, as shown in Fig.~\ref{energy2}(f).

Besides, using $d_z(\mathbf{k})=\pm\frac{1}{2}\{U^{\frac{2}{3}}-
[4(d_x^2(\mathbf{k})+d_y^2(\mathbf{k}))]^{\frac{1}{3}}\}^{\frac{3}{2}}$,
III-type degenerate points can be found at the inner and outer edges of cone structure and tubed structure. In particular, the nonlinear strength $U$ and topological parameter $u$ that makes the Bloch band gaps closed can be given  by the III-type degenerate points. And tubed structure in the excited Bloch band  and cone structure in the ground Bloch band merge gradually into one continued Bloch band without gap, as shown in Fig.~\ref{energy2}(f), which we will give details in the next section by the virtue of effective nonlinear Hamiltonian.

To simplify the discussion,  although the cone structures and tubed structures form in two-dimensional momentum space,  we will use the one-dimensional view of these structures to explain dynamics of nonlinear systems and nonlinear Bloch band in the following section. Thus, cone structures and tubed structures in  one-dimension is of loop.

\subsection{Analysis on gap closing  in trivial and nontrivial topological regime}
In this section, we discuss gap closing at different values of parameters via III-type degenerate point. Linear Hamiltonian have an energy gap in the insulator regime. With tuning the parameter, energy gap can be closed and reopened. Based on Bloch band structures in nonlinear system above, gap between different Bloch bands with increasing $U$ in topological nontrivial regime, as shown in Fig.~\ref{energy2}(f)-(h), can be closed and cannot be reopened anymore, while gap in topological trivial regime remain opened  with increasing $U$, as shown in Fig.~\ref{energy2}(e).
In contrast to gap closing in linear Hamiltonian, here we show the gap closing between different Bloch bands in nonlinear systems by tuning  nonlinear strength $U$ and topological parameter $u$.
For the sake of concise expression, here we use the effective Hamiltonian of  the nonlinear Chern insulator model(Eq.~(\ref{nlchern})) in the vicinity of $\mathbf{k}=(\pi,\pi)$ to illustrate the gap
\begin{eqnarray}
\hat{H}_{eff}&=&(-p_x)\sigma_x+(-p_y)\sigma_y + p_z\sigma_z\nonumber\\
&+&U[|\phi_1(\mathbf{p})|^2,0;0,|\phi_2(\mathbf{p})|^2],\nonumber\\
\end{eqnarray}
where $p_z=u-2+\frac{1}{2}(p_x^2+p_y^2)$, $p_x=k_x-\pi$, and $p_y=k_y-\pi$, then we get $\mathbf{p}$-dependent algebraic eigenvalue equation for $\epsilon$ via the effective Hamiltonian
\begin{eqnarray}\label{}
(\epsilon(\mathbf{p})-U)^2[(\epsilon(\mathbf{p})-U/2)^2 &-& p_x^2-p_y^2] \nonumber\\
&-& p_z^2(\epsilon(\mathbf{p})-U/2)^2=0.\nonumber\\
\end{eqnarray}
Note that different Bloch band structures would be given by different topological parameter $u$ and nonlinear strength $U$.

\begin{figure*}
\includegraphics[width=0.9\textwidth]{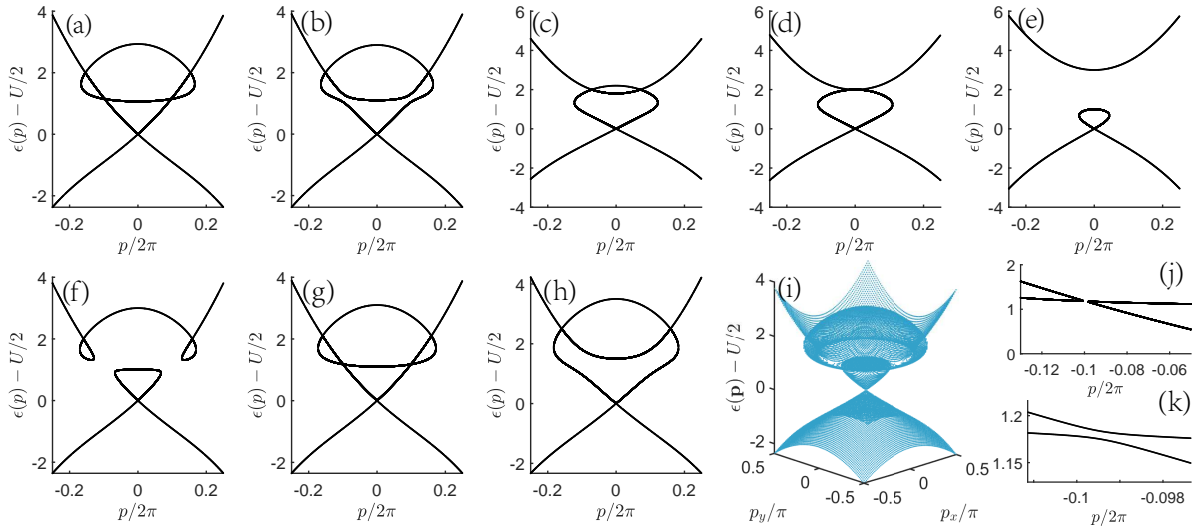}
\caption{Bloch band structures in effective Hamiltonian with different topological parameter $u$ and nonlinear strength $U$, and fixed $p_x=p_y=p$. In the upper panel we give the structures based on increasing $u$ and fixed $U=4$ in (a) $u= 1.066$, (b) $u=1.1$, (c) $u=1.8$, (d) $u=2$, and (e) $u=3$. In the lower panel we give the structures with increasing $U$ (f) $U=4$, (g) $U=4.2$, and (h) $U=5$ where we use fixed $u=1$, and meanwhile we give gap closing point between nonlinear strength (j) $U=4.1993$ and (k) $U=4.1997$. And we give Bloch bands with varied $p_x$ and $p_y$ in (i)$u=1,U=4$.}\label{closingenergy}
\end{figure*}

In Fig.~\ref{closingenergy}, we show the Bloch band structures calculated by effective Hamiltonian. In contrast to original nonlinear Bloch Hamiltonian, Bloch band obtained by the effective Hamiltonian would  be more simple, and tubed structure would  appear  around the circle $p_x^2+p_y^2=4-2u.$  This  structure is clearly shown in Fig.~\ref{closingenergy}(i), which can be produced by rotating the Fig.~\ref{closingenergy}(f) around  the axis of $p=0$. Considering the topological nontrivial regime $u=1$ with nonlinear strength $U=4$ as shown in Fig.~\ref{closingenergy}(f), the Bloch band would produce looped structure in the excited Bloch band and  the ground band jointly. Here we only show the cross-section $p_x=p_y=p$ of Bloch band.

We tune $u$ with fixed $U=4$. When $u\approx 1.066$, gap between different Bloch bands closes, as shown in Fig.~\ref{closingenergy}(a), and the three looped structures in the lower and upper band  merge into a Bloch band. After that, when $u=1.1$, as shown in Fig.~\ref{closingenergy}(b),  these structures form a new large looped structure, and the excited Bloch band and ground Bloch band overlap. Next looped structure deforms with increasing $u$, as shown in Fig.~\ref{closingenergy}(b)-(d). In Fig.~\ref{closingenergy}(d), the looped structure from the ground Bloch band is attached to the excited Bloch band so that gap between them close. After that, the gap starts to reopen, like the gap between Bloch bands with $u=3$ in Fig.~\ref{closingenergy}(e). Here we show the range of the topological parameter $u$ that makes gap closed is $u\approx [1.066,2]$.

Alternatively, we tune $U$ with fixed $u=1$,  where we see the gap opened in topological trivial regime permanently. After $U=4$ in Fig.~\ref{closingenergy}(f), the three looped structures approach to each other, as shown in Fig.~\ref{closingenergy}(g), which is similar to Fig.~\ref{closingenergy}(a). Then gap closing can be observed  between the two enlarged Bloch band structures, as shown in Fig.~\ref{closingenergy}(j)-(k). After that, a new looped structure can be formed[see Fig.~\ref{closingenergy}(h)], which is similar to Fig.~\ref{closingenergy}(b).

At the III-type degenerate point, critical $U$ and $u$ that make gap closed can be found numerically via $d_z(\mathbf{p})=-\frac{1}{2}\{U^{\frac{2}{3}}-[4(d_x^2(\mathbf{p})+d_y^2(\mathbf{p}))]^{\frac{1}{3}}\}^{\frac{3}{2}}$. With the effective Hamiltonian, the algebraic equation can be given by
\begin{eqnarray}
[u-2 + \frac{1}{2}(p_x^2+p_y^2)]=-\frac{1}{2}\{U^{\frac{2}{3}}-[4(p_x^2+p_y^2)]^{\frac{1}{3}}\}^{\frac{3}{2}},
\end{eqnarray}
where $p_x=p_y=p$ is specified. From this equation we can find  that III-type degenerate points satisfy $\epsilon(-p)=\epsilon(p)$.  There are in general  four different $p$ in quasimomentum space and four corresponding III-type twofold degenerate points $\epsilon(p)$. In particular, only two different $p$ and two corresponding III-type twofold degenerate points can be given numerically by careful  choices of $U$ and $u$. This indicates that  two  III-type twofold degenerate points merge into a III-type twofold degenerate points so that gap between different bands disappear, see Fig.~\ref{closingenergy}(a). Thus critical $U$ and $u$  for gap closing can be specified numerically. In Fig.~\ref{closingenergy}(a), two different $p$ and two corresponding III-type twofold degenerate points can be given numerically with $u \approx 1.066$ and $U=4$. In Fig.~\ref{closingenergy}(g),(j)-(k),  we find that two different $p$ and two corresponding III-type twofold degenerate points can be given numerically when $u=1$ and $U_g\approx 4.1996$, falling exactly in the range $U_g\in[4.1993,4.1997]$ (see Fig.~\ref{closingenergy}(j)-(k)). In addition,   manipulating  $u$, we can also tune additional parameter $U$ to close the gap.

\section{Linear response of nonlinear Chern insulator to external field}\label{sec4}
\begin{figure}
\includegraphics[width=0.49\textwidth]{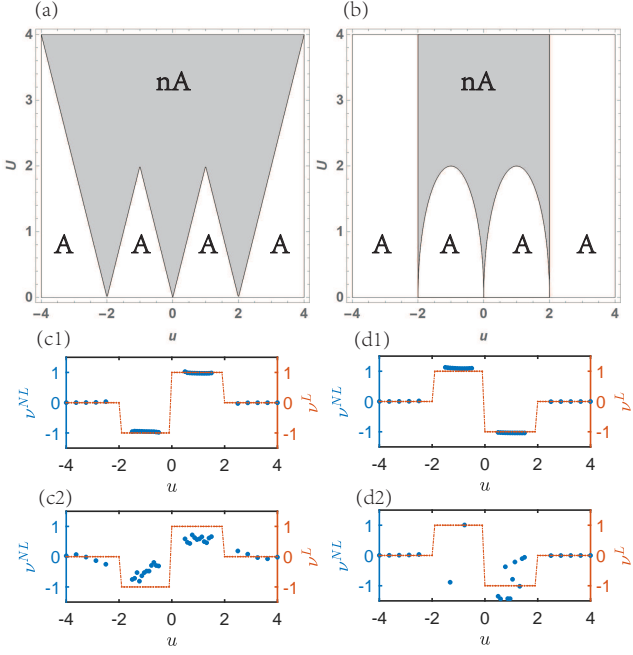}
\caption{Linear response of  nonlinear system to external field $F=0.01$.  (a) and (b) show diagrams with parameter $u$ and $U$ in ground Bloch band and excited Bloch band respectively, where white region denotes adiabatic evolution(A), and gray region denotes non-adiabatic evolution(nA).  Dashed lines in (c1), (c2), (d1) and (d2) denote linear response of linear Chern insulator, and $\nu^L$ on the right side denotes quantized number. Blue Points denote linear response in nonlinear Chern insulator, and $\nu^{NL}$ on the left side denote nonlinear modified quantized number. At last, nonlinear strength $U$ are $U=0.5$ in (c1) and (d1), and $U=3$ in (c2) and (d2). }\label{Nlres}
\end{figure}

To investigate the linear response of  nonlinear Chern insulator, we examine the dynamics of the nonlinear Chern insulator discussed above driven by external field $F$. We first consider the time-dependent nonlinear Bloch Hamiltonian $\hat{H}(t)=H(\mathbf{k}(t))$, the corresponding wave vector given by $\mathbf{k}=[k_x,k_y(t)]$ and  external field $F$ can be introduced by $k_y(t)=k_{y0}+Ft$. Initially, consider the nonlinear eigenstate $|\Psi_{n}^{NL}(t)\rangle$ at $t=0$ in a given Bloch band, and then the expectation of non-vanishing velocity $\hat{v}_x(k_x,k_y(t))=\frac{1}{\hbar}\partial_{k_x} \hat{H}(t)$ with time-dependent density matrix can be calculated numerically by
\begin{eqnarray}
v_n^x(k_x,k_y(t))&=&\langle\Psi(t)|\hat{v}_x(k_x,k_y(t))|\Psi(t)\rangle\nonumber\\
&=&\frac{1}{\hbar}tr[\hat{\rho}(t)\partial_{k_x} \hat{H}(t)],
\end{eqnarray}
where $n=1,2$ denotes ground Bloch band or excited Bloch band and $\hat{\rho}(t)$ denotes the time-dependent density operator. Then we use $k_y(t)=k_{y0}+Ft$ to obtain $\partial_t = F\partial_{k_y}$, and then we consider ground Bloch band and excited Bloch band respectively. The current in a complete Bloch band is given by
\begin{eqnarray}
J_n^x &=& \sum_{k_x,k_y} v_n^x(k_x,k_y(t)).
\end{eqnarray}
And the current\cite{Shen2012} is given by
\begin{eqnarray}
J_n^x=  \sigma_{xy} F,
\end{eqnarray}
where $\sigma_{xy}$ is the linear response of this nonlinear Chern insulator
\begin{eqnarray}
\sigma_{xy} =  \frac{1}{F} \sum_{k_x,k_y}   v_n^x(k_x,k_y(t)) =\frac{\nu^{NL}}{h},
\end{eqnarray}
where $\nu^{NL}$ is a quantized number modified by nonlinearity, which imply that quantized response of nonlinear systems could be broken. Consider the Bloch band structures of nonlinear Chern insulator in Sec.\ref{sec3}, we show the diagram with different parameters $u$ and $U$, based on whether cone structures and tubed structures appear. In Fig.~\ref{Nlres}(a) and (b), there are two regions, one of which is adiabatic region(A) and the other is non-adiabatic(nA) region, which are linked to adiabaticity of nonlinear systems. In linear regime, we give a quantized response with different parameter $u$, as shown by the red line in Fig.~\ref{Nlres}. Nevertheless, in nonlinear regime, combining linear response with the nonlinear Bloch band structures, there are two kinds of linear response, one of which is non-quantized response modified by nonlinearity, as shown in Fig.~\ref{Nlres}(c1) and (d1), while the other type of the linear response can be broken completely and generate the drastic change deviate from quantized response, as shown in Fig.~\ref{Nlres}(c2) and (d2). And then these two different kinds of destruction of quantized linear response correspond to the two regions of the diagram in Fig.~\ref{Nlres}(a) and (b). What happens to linear response of nonlinear systems to external field?  Because adiabatic evolution in nonlinear systems is broken due to the appearance of new structures.  This question can be tackled and shown using numerical simulations of dynamics of nonlinear systems. In the following section, for the sake of simplicity, we will choose one of the adiabatic evolution paths to show the dynamics of nonlinear systems and the failure of adiabatic evolution due to the new structure.

\section{Adiabatic evolution of nonlinear Chern insulator}\label{sec5}
As a result of the appearance  of the new  structures, dynamics of nonlinear systems is not trivial anymore. In the previous studies, authors shown  that the adiabatic evolution would be broken when the system arrives at the tail of loop structure\cite{niu2000,niu2002,blume2020}. In this section, we explore this problem again but for our topological systmes.
For simplicity, we start with considering an evolution path in $k_x=k_y$ direction.  Considering that $\mathbf{k}$ in quasimomentum increases from $(k_x,k_y)=(0,0)$ to $(k_x,k_y)=(2\pi,2\pi)$,  time evolution state is initially  prepared in the eigenstate of nonlinear Hamiltonian at quasimomentum  $(k_x,k_y)=(0,0)$ and then evolves governed  by Eq.~(\ref{timenl}) with $\mathbf{k}=\mathbf{F}t+\mathbf{k}_0$ and $\mathbf{k}_0=0$, where a very weak constant acceleration force $\mathbf{F}$ is given by $\mathbf{F}=(F_x,F_y)=(F,F)$. This force  is equivalent to external field in the last section.


\subsection{dynamics of  nonlinear Chern insulator}
\begin{figure}
\includegraphics[width=0.49\textwidth]{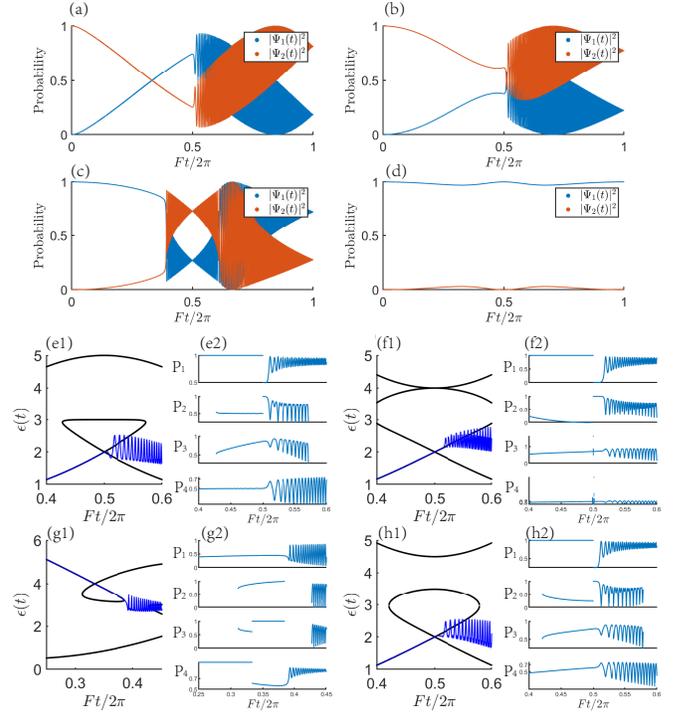}
\caption{Time evolution, dynamical energy level, and transition probability between the resulting  state and the instantaneous eigenstates with different parameters and $k_x=k_y=k=Ft$. To calculate the time evolution, we set acceleration force $F=0.01$ and nonlinear strength $U=4$. In (a)-(d), time-dependent probability with different initial states, eigenstate at $k=0$ in the ground Bloch band is set as the initial state in (a) and (c), and eigenstate at $k=0$ in the excited Bloch band is set as the initial state in (b) and (d). The other  parameters chosen  are (a) $u=1$, (b) $u=1$, (c) $u=2.5$, and (d) $u=2.5$.  Here we show four different Bloch band structures and the corresponding transition probability in (e1) $u=1$, (f1) $u=2$, (g1) $u=1$ and (h1) $u=2.5$. In (e1), (f1), and (h1), initial state is eigenstate at $k=0$ in the ground Bloch band, and initial state is eigenstate at $k=0$ in the excited Bloch band in (g1). For comparison, the local features of complete evolution process are also shown. }\label{dynamicalevel}
\end{figure}

Here we consider acceleration force $F_x=F_y=F$. Along the path, looped structures can be obtained from the nonlinear Bloch band structures. We address the  probability $|\Psi_i(t)|^2, i=1,2$ obtained from the time evolution state $\Psi(t)=[\Psi_1(t),\Psi_2(t)]^T$. Firstly, consider the topological nontrivial regime, the probability $|\Psi_i(t)|^2, i=1,2$  evolves adiabatically from instantaneous eigenstate of the ground Bloch band at $k=0$, and then there is an irregular oscillation around the I-type degenerate point due to looped structure in the ground Bloch band, as shown in Fig.~\ref{dynamicalevel}(a) and (e1). Then using eigenstate from the excited Bloch band at $k=0$ as an initial state, we show the time evolution in Fig.~\ref{dynamicalevel}(c), and band structure in Fig.~\ref{dynamicalevel}(g1). We can observed that  the excited Bloch band produce a looped structure, and there is  an irregular oscillation at the tail of looped structure from the excited Bloch band.

In topological trivial regime, one looped structure is produced at the ground Bloch band, as shown in Fig.~\ref{dynamicalevel}(b), (d) and (h1). And the initial state in Fig.~\ref{dynamicalevel}(b) and (d) is set to be the  eigenstate from ground Bloch band at $k=0$ and the eigenstate from excited Bloch band at $k=0$, respectively. Here irregular oscillation in the ground Bloch band appear around  the I-type degenerate point, as shown in Fig.~\ref{dynamicalevel}(b). Whereas in the excited Bloch band, looped structure cannot be produced so that there is not irregular oscillation along the path. This indicates that adiabatic evolution is satisfied, as shown in Fig.~\ref{dynamicalevel}(d).

In short, the presence of nonlinearity induced novel structures alter the time evolution dynamics and the adiabaticity breaks down in the  nonlinear Chern insulator. This leads to  non-adiabatic time evolutions and consequently  the  linear response to external fields changes drastically, as shown in Fig.~\ref{Nlres}(c2) and (d2). On the contrary, the adiabatic evolution would give  quantized linear responses, as shown in Fig.~\ref{Nlres}(c1) and (d1).

\subsection{average  energy and adiabaticity}
In order to calculate the expectation value of the time-dependent nonlinear Hamiltonian, we  compute  numerically the wave function at time $t$   by solving $i\hbar\partial_t\Psi(t)=\hat{H}(t)\Psi(t)$. Substituting the wave function $|\Psi(t)\rangle$ into, $\epsilon(t)=\langle\Psi(t)|\hat{H}(t)|\Psi(t)\rangle$, the expectation of the Hamiltonian will be given. With different  parameters, we find different structures (see Fig.~\ref{dynamicalevel}(e1)-(h1)) and different time evolution of average $\epsilon(t)$.
When $u<2$, namely in nontrivial regime, $\epsilon(t)$ starts oscillating  around the I-type degenerate point as shown in Fig.~\ref{dynamicalevel}(e1)-(e2),  while it starts oscillating  at the tail of the looped structure in the excited Bloch band, see Fig.\ref{dynamicalevel}(g1). The simulation was performed with the same initial states as in the last section. We also find that
when $u>2$ (see Fig.~\ref{dynamicalevel}(h1)), irregular oscillations in $\epsilon(t)$  start around the I-type degenerate points, this is  the same as $\epsilon(t)$ in the nontrivial regime that gets start  from the ground instantaneous eigenstate at $k=0$.
When $u=2$ (see Fig.~\ref{dynamicalevel}(f1)), we find that irregular oscillations in $\epsilon(t)$ start behind the appearance of degenerate point. Namely, the oscillations start at a value larger than that at which the degenerate point occurs.
In other words, irregular oscillations appear around the degenerate point, not exactly at the point. In particular, at the critical point $u=2$, the distance between the location of irregular oscillation and  the degenerate point  is  the largest one.

In the following, we would use the populations on each  instantaneous bands  to quantify  the adiabaticity.
To obtain the transition probability, a set of  $\mathbf{k}$-dependent  eigenstates $\chi(\mathbf{k})$ of nonlinear Bloch Hamiltonian can be given by Eq.~(\ref{eigennl}) and then instantaneous eigenstates $\chi_i(\mathbf{k}(t))$ is obtained. And transition probability can be define as, $P_i=|\chi_i(\mathbf{k}(t))^\dagger\Psi(\mathbf{k}(t))|^2$. Generally speaking, these nonlinear instantaneous eigenstates cannot be orthogonality and sum of these transition probability will not be unity.  There are some details on instantannoues eigenstates in Appendix.~\ref{app}.
Using the transition probability $P_i=|\chi_i^\dagger(\mathbf{k})\Psi(\mathbf{k}(t))|^2$, the difference between evolution state and instantaneous eigenstates are shown in Fig.~\ref{closingenergy}(e2)-(h2). The transition probability  $P_1-P_4$ correspond to  the instantaneous eigenstates with eigenvalues from small to large. When $P_i$ is unity which imply that evolution state is the same as  nonlinear instantaneous eigenstates, adiabatic evolution is satisfied, and while  transition probability is irregular oscillation which implies that these two states are different from each other, adiabatic evolution is broken.

\section{Conclusions}\label{sec6}
In summary,  the Bloch band structures and gap closing  has been discussed for generalized  Chern insulator described by a nonlinear Hamiltonian.  In contrast to linear systems,  Bloch band in nonlinear systems possesses  additional structures around degenerate points determined by the competition between the nonlinearity and other parameters of the system. The observed features are completely different from that in linear systems. For example, in nonlinear Chern insulator, cone structure from the ground Bloch band and tubed structure from the excited Bloch band appear.
From the other aspect, the  linear response of the system to external fields is also different from the linear one. Namely, the response is not quantized due to nonlinearity.
To explain this feature, we examine  the adiabaticity of the time evolution and find that  in non-adiabatic  regime,  irregular oscillations   appear around the center of the cone   or the tail of the tube of the energy bands.
Experimentally, band structures and tunneling probabilities in nonlinear systems can be observed  in ultracold atoms in a moving one-dimensional optical lattice\cite{blume2020}. Alternatively, it is also realizable in  nonlinear circuit arrays, which can be constructed  by setting one-dimensional transmission-line circuits in electronic systems\cite{Hadad2018}. Strong nonlinear regime can be realized in   a one-dimensional mechanical Su-Schrieffer-Heeger interface model with nonlinearity\cite{alexander2021}.
Finally, our method to describe Bloch band structures in nonlinear systems can be extended to   other nonlinear topological systems, for example  high dimensional systems and systems with different nonlinear strength.
We conjecture  that pursuing these interesting band structures in topological systems  with nonlinearity would  be the main concern in  future works.

\begin{acknowledgments}
We thank Hongzhi Shen and Dexiu Qiu for helpful discussions. This research was funded by National Natural Science Foundation of China (NSFC) under
Grants No. 12175033 and No. 12147206,  and National Key R$\&$D Program of China(No.2021YFE0193500).
\end{acknowledgments}

\appendix

\section{Eigenstates of nonlinear Bloch Hamiltonian}\label{app}
\subsection{Non-degenerate regime}
By virtue of nonlinear Bloch Hamiltonian, eigenstates can be obtained.  Here we assume that these eigenvectors are normalized, but generally speaking, they are not orthogonal to each other due to nonlinearity. Nonlinear Bloch Hamiltonian is given by
\begin{eqnarray}
H(\mathbf{k})=\frac{U}{2}\sigma_0+d_x(\mathbf{k})\sigma_x+d_y(\mathbf{k})\sigma_y+d_z^{NL}(\mathbf{k})\sigma_z
\end{eqnarray}
where $d_z^{NL}(\mathbf{k})=d_z(\mathbf{k})(\frac{2\epsilon(\mathbf{k})-U}{2(\epsilon(\mathbf{k})-U)})$, specially, when $U=0$, it will be $d_z^{NL}(\mathbf{k})=d_z(\mathbf{k})$. And eigenstate(omitting $\mathbf{k}$ for simplicity) is given by
\begin{eqnarray}\label{gnl}
\chi=
\frac{1}{\sqrt{2(\epsilon-\frac{U}{2})(\epsilon-\frac{U}{2}-d_z^{NL})}}
\left(
\begin{array}{c}
d_x-id_y \\
\epsilon-\frac{U}{2}-d_z^{NL}
\end{array}\right),\nonumber\\
\end{eqnarray}
where $\chi=[\chi_1,\chi_2]^T$, and we get general eigenstates in nonlinear systems, but we need to make detailed discussion about the eigenstates of  degenerate eigenvalues.

\subsection{Degenerate regime}
There are three kinds of eigenstates for the three types of degenerate points. Consider $d_x^2+d_y^2=0,\epsilon=U/2$, namely, I-type degenerate point, and use the expression of $|\chi_1|^2,|\chi_2|^2$ given in the main manuscript, then  substitute $\epsilon$ into $|\chi_1|^2,|\chi_2|^2$, and eigenstates of the I-type degenerate point is given by
\begin{eqnarray}\label{gdl}
\chi_{d_x^2+d_y^2=0}=
\left(
\begin{array}{c}
\sqrt{\frac{1}{2}+\frac{d_z}{2(\epsilon-U)}}e^{i\theta} \\
\pm\sqrt{\frac{1}{2}-\frac{d_z}{2(\epsilon-U)}}
\end{array}
\right),\nonumber\\
\end{eqnarray}
where $\theta$ are arbitrary phase depending on which direction we approach the degenerate point.

Consider $d_z=0,\epsilon=U$, namely, II-type degenerate point, $d_z^{NL}$ is given by
\begin{eqnarray}
d^{NL}_z=d_z+\frac{\kappa}{2}
\end{eqnarray}
where $d_z^{NL}$ is $\kappa$-dependent which is considered in the main manuscript, and corresponding $(\kappa,\epsilon)=(\pm\frac{\sqrt{U^2-4(d_x^2+d_y^2)}}{U},U)$, and then substitute $\kappa,\epsilon$ into $d_z^{NL}$, and eigenstates of the II-type degenerate point is given by
\begin{eqnarray}
\chi_{d_z=0}=
\frac{1}{\sqrt{2(\epsilon-\frac{U}{2})(\epsilon-\frac{U}{2}-\frac{\kappa}{2})}}
\left(
\begin{array}{c}
d_x-id_y \\
\epsilon-\frac{U}{2}-\frac{\kappa}{2}
\end{array}
\right).\nonumber\\
\end{eqnarray}

Consider $d_z=\pm\frac{1}{2}\{U^{\frac{2}{3}}-[4(d_2^2+d_y^2)]^{\frac{1}{3}}\}^{\frac{3}{2}}, \epsilon=U/2+1/2[4U(d_x^2+d_y^2)]^{\frac{1}{3}}$, namely, III-type degenerate point, eigenstate is given by
the form of eigenstates in Eq.~\ref{gdl}.



\end{document}